\documentclass{aa}
\usepackage{graphicx}
\usepackage{epsfig}
\usepackage{subfigure}
\usepackage{amsmath}
\usepackage{txfonts}
\def\kms{\, {\rm km}\, {\rm s}^{-1}}
\def\msy{\, {\rm M}_\odot \, {\rm yr}^{-1}}
\def\mdotw{\, \dot{M}_{\rm w}}
\def\pram{\, P_{\rm ram}}
\def\Pb{\, P_{\rm b}}
\def\Pcf{\, P_{\rm cf}}
\def\vw{\, V_{\rm w}}
\def\rhow{\, \rho_{\rm w}}
\def\Lw{\, L_{\rm w}}
\def\Rw{\, R_{\rm w}}
\def\Egrb{\, E_{\rm GRB}}
\def\Mgrb{\, M_{\rm GRB}}

\def\vs{\, V_{\rm s}}
\def\Gams{\, \Gamma_{\rm s}}

\def\Rs{\, R_{\rm s}}
\def\Rst{\,R_{\rm ST}}
\def\Rd{\, R_{\rm d}}

\def\half{\, \frac{1}{2}}
\def\modot{\, {\rm M}_\odot}
\def\gcm{\, {\rm g}\, {\rm cm}^{-3}}

\begin{document}
\title{Forming a constant density medium close to long gamma-ray bursts}

    \author{A. J. van Marle
           \inst{1}
           \and
           N. Langer
           \inst{1}
          \and
           A. Achterberg
           \inst{1}
           \and
           G. Garc{\a'i}a-Segura
           \inst{2}
           }

    \offprints{A. J. van Marle}

    \institute{Astronomical Institute, Utrecht University,
               P.O.Box 80000, 3508 TA, Utrecht, The Netherlands\\
               \email{A.vanMarle@astro.uu.nl}\\
               \email{N.Langer@astro.uu.nl}\\
               \email{A.Achterberg@astro.uu.nl}\\
          \and
              Instituto de Astronom{\a'i}a-UNAM,
              APDO Postal 877, Ensenada, 22800 Baja California, Mexico\\
              \email{ggs@astrosen.unam.mx}
              }

    \date{Received <date> / Accepted <date>}

  \abstract{}
{The progenitor stars of long Gamma-Ray Bursts (GRBs) are thought to be
Wolf-Rayet stars, which generate a massive and energetic wind. Nevertheless,
about 25~percent of all GRB afterglows light curves indicate
a constant density medium close to the exploding star.
We explore various ways to produce this, by creating situations where
the wind termination shock arrives very close to the star, as the shocked
wind material has a nearly constant density. }
{Typically, the distance between a Wolf-Rayet star and the wind termination shock is too large
to allow afterglow formation in the shocked wind material.
Here, we investigate possible causes allowing for a smaller distance:
A high density or a high pressure in the surrounding interstellar medium (ISM),
a weak Wolf-Rayet star wind,
the presence of a binary companion, and
fast motion of the Wolf-Rayet star relative to the ISM.}
{We find that all four scenarios are possible in a limited parameter space,
but that none of them is by itself likely to explain the large fraction
of constant density afterglows. }
{A low GRB progenitor metallicity, and a high GRB energy make the 
occurrence of a GRB afterglow in a constant density medium more likely.
This may be consistent with constant densities being
preferentially found for energetic, high redshift GRBs.}


   \titlerunning{Constant density medium close to a GRB progenitor}
   \authorrunning{van Marle et al.}
   \maketitle
%
%
%

\section{Introduction}

\begin{figure}[!t]
    \centering
    \resizebox{\hsize}{!}{\includegraphics[width=0.95\textwidth]{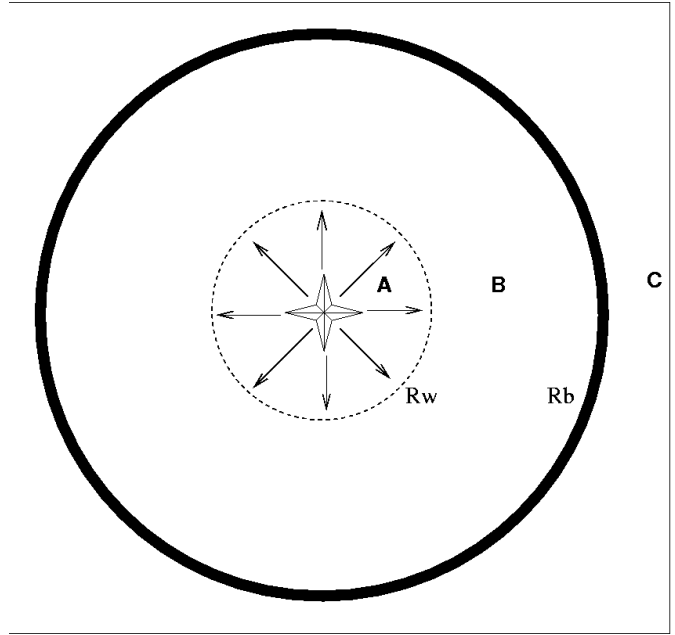}}
       \caption{Schematic view of a circumstellar bubble (not to scale).
       The free-streaming stellar wind (region A) passes through the wind termination shock $\Rw$ (dashed line) to enter the hot bubble of shocked
       wind material (region B).  The high thermal pressure in the hot bubble sweeps up a shell (thick, continuous line) of radius $R_{\rm b}$, which expands into
       the ISM (region C).
        }
          \label{fig:bubble}
    \end{figure}

  During their evolution, massive stars lose a major fraction of their mass in the form of  a stellar wind.
   The interaction between this stellar wind and the surrounding interstellar medium creates a circumstellar bubble.
   Within approximations, the morphology and evolution of such  bubbles can be predicted analytically
  (Castor et al. \cite{CMW75}; Weaver et al. \cite{WCM77}; Ostriker \& McKee \cite{OM88}), or by numerical simulations
  (Garc{\a'i}a-Segura et al. \cite{GML96}, \cite{GLM96}; Freyer et al. \cite{FHY03}; van Marle et al. \cite{MLG04}, \cite{MLG05a}, \cite{MLG05b}).
   Figure \ref{fig:bubble} shows schematically the morphology of a spherical wind bubble.
   Closest to the star is the free-streaming stellar wind.
   Its material hits the wind termination shock, where the kinetic energy of the wind material is thermalized,
   which feeds the 'hot bubble' of shocked wind material. Outside the hot bubble lies a shell of shocked
   interstellar gas that has passed through
   the outer shock that delimits the bubble as long as it expands supersonically with respect to the interstellar medium (ISM).
   The material forming this shell is swept up as the high-pressure gas of the hot bubble.
   Finally, there is the ambient medium, which has not yet been overrun by the expanding circumstellar bubble.

   Within the collapsar picture (Woosley \cite{W93}), Wolf-Rayet stars, which form the final evolutionary
   stage of massive stars, are thought to be the progenitors of long Gamma-Ray Bursts (GRBs).
   This relativistic jet associated with the GRB expands into the circumstellar medium (CSM).
   On its way it first encounters the free-streaming wind of the Wolf-Rayet star,
   and only afterwards it may penetrate into the shocked wind material.

   The medium close to a Wolf-Rayet star has the density profile of a free-streaming wind with a total mass flux $\dot{M}$
   and velocity $V_{\rm w}$:
\begin{equation}
\label{eq:fwind}
         \rhow = \frac{\mdotw}{4 \pi r^{2} \vw} \; .
\end{equation}
  The wind density drops as $\rho_{\rm w} \propto r^{-2}$ once the wind has reached its terminal velocity,
  which occurs at a distance of a few stellar radii.
  Both analytical and numerical calculations suggest that the free wind of a Wolf-Rayet star 
  typically extends over a radius of the order of a few parsec.
  This is confirmed by observations of Wolf-Rayet nebulae, such as \object{NGC~6888} and \object{RCW~58}, which also have radii of
  the order of several pc (Gruendl et al. \cite{GCDP00}).

   However, from the analysis of many GRB afterglows, a {\em constant} circumstellar medium density  is inferred.
   The fraction of GRBs that show a constant density profile varies, depending on the afterglow model,
   but it can be estimated to $\geq 25~$percent (Chevalier \& Li \cite{CL00}; Panaitescu \& Kumar \cite{PK01}; Panaitescu \& Kumar, \cite{PK02}; Chevalier et al. \cite{CLF04}).
   Moreover, it is noteworthy that all studies mentioned were limited to GRBs
   detected before the launch of the SWIFT satellite, i.e.
   they are primarily concerned with low redshift GRBs. It cannot be excluded that the fraction
   of afterglows in a constant density medium is higher for larger redshift.

   The density of the CSM and its radial dependence is also relevant for the use of GRBs
   as standard candles in cosmological studies.
   For this purpose, the correlation between peak energy and corrected collimated energy has to be calculated,
   which depends on the nature of the circumstellar or interstellar medium
   (Ghirlanda et al. \cite{Getal05}).

   Here, we investigate various possibilities to obtain GRB afterglows in a constant density medium.
   In Sect. \ref{sec-analyzes} we estimate analytically at what radius the constant
   density medium needs to occur in order to appear
   in the GRB afterglow lightcurve.
   In Sects. \ref{sec-windtermshock} and \ref{sec-combinations} we explore
   ways to produce
   a constant density medium sufficiently close to the star, using both analytical and numerical techniques to model the
   circumstellar medium. In Sect. \ref{sec-conclusion} we discuss the likelihood of the various possibilities.

\section{Analytic considerations}
\label{sec-analyzes}

   While the region closest to the Wolf-Rayet star has a density that decreases with the radius squared, the next layer of the circumstellar bubble
  (region B in Fig. \ref{fig:bubble}) has a nearly constant density, essentially since the high sound speed inside this material
   precludes the occurrence of large pressure- and density-gradients.
   This hot bubble of shocked wind gas drives a shell into the CSM (Weaver et al. \cite{WCM77}).
   It has been proposed (Wijers et al. \cite{W01}; Chevalier et al. \cite{CLF04}) that the GRB afterglows are created while the jet moves through this medium.
   This requires the Gamma-Ray Burst jet to cross the wind termination shock at an early stage.
   Since Wolf-Rayet stars have a strong wind, with a high mass loss rate and a high terminal velocity, the ram pressure of the wind is
   considerable, and
   the wind termination shock that separates this bubble from the free-streaming wind is typically located at a distance
   of 10~pc from the star, which would make this scenario impossible.
   (For a discussion of the models by Chevalier et al. (\cite{CLF04}), which find considerably smaller shock radii, 
   see Sect. 3.3).

   \subsection{The location of the wind termination shock}
   \label{sec-reverseshock}

   The location of the wind termination shock can be calculated analytically, using the so-called thin-shell approximation
  (Castor et al. \cite{CMW75}; Weaver et al. \cite{WCM77}; Ostriker \& McKee \cite{OM88}).
   This assumes that the forward shock is completely radiative, whereas the hot bubble of shocked wind material is adiabatic, which                
   serves as a good first approximation.
   The mechanical luminosity of the star,
   \begin{equation}
   \label{eq:mechlum}
      \Lw = \half \mdotw \vw^2 \; ,
   \end{equation}
   is a measure of the mechanical energy that the star puts into the CSM.
   For a given mechanical luminosity, it is possible to calculate the thermal pressure in the bubble as a function of time,
   by balancing the pressure in the hot, shocked wind with the ram pressure $\rho_{0} ({\rm d} R_{\rm b}/{\rm d}t)^2$ due to the
   expansion into a constant-density ISM with mass density $\rho_{0}$ (Weaver et al. \cite{WCM77}):
  \begin{equation}
  \label{eq:Pbubble}
       \Pb = \frac{7}{(3850 \pi)^{2/5}} \Lw^{2/5} \rho_0^{3/5} t^{-4/5}.
   \end{equation}
   This ansatz assumes that the ISM is cold, so that it has no significant thermal pressure.
   Since the mechanical luminosity of the star can change considerably over time, we have to take the average value for the
   mechanical luminosity.

   The wind termination shock is located at the radius $R_{\rm w}$ such that the thermal pressure in the shocked wind material
   equals the hot bubble pressure $P_{\rm b}$. The post-termination shock pressure is roughly equal to the ram pressure in the wind,
   \begin{equation}
   \label{eq:Pram}
    \pram = \rhow \vw^2 = \biggl(\frac{\mdotw \vw}{4 \pi r^2} \biggr) \; .
   \end{equation}
   Combining Eqs.~\ref{eq:Pbubble} and \ref{eq:Pram} gives the radius of the wind termination shock:
   \begin{equation}
   \begin{aligned}
   \label{eq:Rshock1}
      \Rw &= \biggl(\frac{\mdotw \vw}{4 \pi \Pb}\biggr)^{1/2} \\
          &= \biggl(\frac{\mdotw \vw}{28}\biggr)^{1/2} \:
        \biggl( \frac{3850}{\overline{\Lw}}\biggr)^{1/5} \:\frac{ t^{2/5}}{(\pi \rho_0)^{3/10}}.
   \end{aligned}
   \end{equation}
   Here the mass loss rate ($\mdotw$) and wind velocity ($\vw$) are time dependent, while $\overline{\Lw}$ is the average mechanical luminosity over the
   entire evolution of the star up to that moment. 
   To obtain the radius of the wind termination at the end of the Wolf-Rayet phase, one should take the properties of the Wolf-Rayet wind for
   $\mdotw$ and $\vw$, and one should use the average value of the mechanical luminocity.\footnote{N.B. This is only true for the situation where the Wolf-Rayet wind expands directly into the main sequence bubble. 
   During the early Wolf-Rayet stage, the Wolf-Rayet wind sweeps up the wind of the preceding phase (Red Supergiant or Luminous Blue Variable). 
   In that case, the radius of the wind termination shock can be found by using Eqn. (31) from Garc{\a'i}a-Segura and Mac-Low (\cite{GM95}).}

   If the ram pressure of the wind changes, the radius of the wind termination shock will change on the hydrodynamical
   time scale of the hot bubble. Since the temperature in the hot bubble is of the order of $10^7$ K,
   this timescale is 1...4$\times 10^4$ years, for typical bubble radii
   of few tens of parsecs (Garc{\a'i}a-Segura et al. \cite{GML96}, \cite{GLM96};
   Freyer et al. \cite{FHY03}; van Marle et al. \cite{MLG05b}).
   Consequently, the wind termination shock will closely follow any changes of the stellar wind parameters
   which occur on the time scale of the stellar evolution.
   The time dependence of the thermal pressure in the bubble is not important here,
   as we only consider the Wolf-Rayet period in the following, which lasts only a fraction
   of the total stellar lifespan.

   \subsection{Afterglow production in a free wind}
   \label{sec-blastwave}

   In order to establish a maximum radius for the wind termination shock in those cases where a constant-density circumstellar medium is inferred,
   we must first calculate the typical distance of the jet from the progenitor star when the afterglow is produced.
   The afterglow of a GRB is believed to be synchrotron radiation, produced when the jet produces a relativistic shock in the surrounding medium.
   For a constant-density medium to show up in the afterglow light curve, the jet must still be relativistic when it passes through
   the wind termination shock.
   
   We make an analytical estimate of the distance that the GRB jet penetrates into the surrounding medium before slowing down to
   non-relativistic speed. C.f. Piran (\cite{P99}, \cite{PI}), Panaitescu \& Kumar (\cite{PK00}) and Chevalier and Lee (\cite{CL00}).
   This provides a rough estimate of the radius at which the constant density medium must be found. 
   For a more accurate result it is necessary to model the GRB afterglow lightcurve generated in a structured circumstellar medium. 
   Such models were presented by Ramirez-Ruiz et al. (\cite{RGSP05}) and Eldridge et al. (\cite{EGDM05}).

    We consider an extremely powerful fireball with an explosion energy $\Egrb~\simeq 10^{52}...10^{54}~{\rm erg}$, and  with a small mass loading
    in the form of baryons: $\Mgrb({\rm baryonic})~\leq 10^{-4}...10^{-5}~\modot$.
    Note that $\Egrb$ is the energy of the fireball for a spherical explosion.
    The actual energy of the jet can be much lower, depending on the opening angle.
    A jet confined to a solid angle $\Omega_{\rm j}$ carries an energy $E_{\rm j}~\sim~\Egrb~\times~(\Omega_{\rm j}/4 \pi)$ and a mass
    $M_{\rm j}~\sim~\Mgrb~\times~(\Omega_{\rm j}/4 \pi)$.

    As the total explosion energy is significantly larger than the rest energy of the ejected mass,
    \begin{equation}
    \Egrb \gg \Mgrb c^2,
    \end{equation}
    the material in the blast wave (or jet) will expand rapidly, moving with a speed $\vs \approx c$ and with a corresponding bulk Lorentz-factor:
    {\bf $\Gams \gg 1$.}
    During the initial phase of the expansion, the explosion energy $\Egrb$ is converted by pressure forces into the bulk motion of the material.
    Once the initial energy released by the GRB has been converted into bulk kinetic energy, the Lorentz-factor is determined
    solely by the ratio between the explosion energy and the rest-mass energy (see the review by Piran, \cite{PI}, and references therein):
    \begin{equation}
    \Gams  \approx \eta \equiv \frac{\Egrb}{\Mgrb c^2}.
    \end{equation}
    This relation will hold until the jet has swept up a sufficient amount of mass (see below), or until (in the case of a jet) sideways expansion
    becomes important.
    In order to explain the observations, we need $\eta~\simeq~100...1000$ (Thompson \cite{T94}).

    During the expansion, the blast wave sweeps up material which passes through a relativistic shock.
    The total energy of the system, consisting of the original mass and radiation liberated at the GRB and the swept-up mass,
    approximately equals
    \begin{equation}
    \label{Etot}
    \Egrb \approx \Gams \: \Mgrb c^2 + (\Gams^2 -1) \: M_{\rm sw} c^2.
    \end{equation}
    The second term is the total (internal) energy of the shocked, swept-up mass. For $\Gams \gg 1$ this term follows from the jump conditions for
    an ultra-relativistic shock (Blandford \& McKee, \cite{BM76}) which give the post-shock energy density as $e_{2} \sim \Gams^2 \rho_{1} c^{2}$ where the
    subscript 1 (2) refers to the pre-shock (post-shock) state, and where $\rho_{1}$ is the rest mass density in front of the shock.
    For a non-relativistic shock with velocity $V_{\rm s}$ the classical Rankine-Hugoniot relations give $e_{2} \approx \rho_{1} V_{\rm s}^{2}$.
    The second term in expression (\ref{Etot}) correctly interpolates between these two limiting cases, up to factors of order unity.
    $E_{\rm GRB}$ is conserved as long as radiation losses can be neglected.

    In the limit $\Gams \gg 1$ one can write:
    \begin{equation}
    \Egrb = \Gams \: \Mgrb c^2 + \Gams^2 \: M_{\rm sw} c^2.
    \end{equation}
    The free expansion phase, where $\Gams \sim \Egrb/ \Mgrb \:c^2 \sim$ constant, ends when the energy contained in the shocked, swept-up mass
    becomes comparable with the energy of the original explosion:
    \begin{equation}
    \label{dcmass}
    M_{\rm sw} \simeq \frac{\Mgrb}{\Gams} \simeq \frac{\Mgrb}{\eta} = \frac{\Egrb}{\eta^2 c^2} \equiv M_{\rm d},
    \end{equation}
    which defines the deceleration mass $M_{\rm d}$. The blast wave or jet will decelerate when $M_{\rm sw} > M_{\rm d}$.
    Taking the simple case of a spherical blast wave expanding into a freely expanding stellar wind, with a density profile as in Eqn. \ref{eq:fwind}
    and constant $V_{\rm w}$, the swept-up mass when the fireball has expanded to a radius $\Rs$ equals
    \begin{equation}
    \label{fwmass}
    M_{\rm sw} = \frac{\mdotw \: R_{\rm s}}{\vw}.
    \end{equation}
    Combining relations (\ref{dcmass}) and (\ref{fwmass}) one finds the {\em deceleration radius}, the radius where the free expansion
    of the blast wave stops, provided this occurs while still in the free-wind region:
    \begin{equation}
    \Rd = \frac{M_d \vw}{\mdotw} \simeq \frac{\Egrb \vw}{\eta^2 \: \mdotw c^2}.
    \label{eq:Rd}
    \end{equation}

    The blast wave now enters the relativistic equivalent of the well-known decelerating Sedov-Taylor expansion, until the
    expansion speed becomes sub-relativistic. After that, it enters the ordinary Sedov-Taylor expansion phase until radiation
    losses become important, invalidating the assumption of the conservation of the total energy (Eqn.\ref{Etot}).
    During the Sedov-Taylor phase, the swept-up mass dominates the energy equation, so that:
    \begin{equation}
    \Egrb = \Gams^2 M_{\rm sw} \: c^2 \; .
    \end{equation}
    As long as the blast wave stays in the free wind, relation (\ref{fwmass}) remains valid, and one has:
    \begin{equation}
    \Gams \approx \sqrt{\frac{\Egrb \vw}{\mdotw c^2 \: R_{\rm s}}} = \left( {\frac{\Rs}{\Rd}} \right)^{-1/2} \; ,
   \end{equation}
    where we have used Eqn. \ref{eq:Rd}.
    We can take the point where $\Gams \approx 1$ as the point where the blast wave becomes non-relativistic,
    This happens at the so-called {\em Sedov-Taylor radius} $\Rst$, which in a free wind equals:
    \begin{equation}
    \label{eq:ST}
    \Rst \equiv \eta^2 \Rd = \frac{\Egrb \vw}{\mdotw c^2} \; .
    \end{equation}
    For $R_{\rm s} > \Rst$ one has from (\ref{Etot}) with $\Gams^2 - 1 \approx V_{\rm s}^{2}/c^{2}$:
    \begin{equation}
        \frac{V_{\rm s}}{c} \sim \left( \frac{R_{\rm s}}{R_{\rm ST}} \right)^{-1/2} \
    \end{equation}
    for as long as the expansion proceeds in the free-wind region ($\Rs < R_{\rm w}$).

    For a typical Wolf-Rayet wind one has: $\Egrb = 5.0\times 10^{52}$~erg, $\vw$ = 2000~$\kms$ and $\mdotw = 3.0\times 10^{-5}~\msy$.
    This gives us $\Rst \sim 1.9$ pc, well within the free-wind region of the circumstellar bubble.
    This is the radius at which the production of the afterglow radiation is expected to end.
    In order to have a constant density medium during afterglow production, the wind termination shock has to lie much
    closer to the star, typically at $\sim 0.1$ pc (Chevalier et al. \cite{CLF04}).
    If the entire afterglow shows a constant density medium, the jet has to pass the wind termination shock before it starts to decelerate.
    i.e. before the end of the free expansion phase.
    The transition from free expansion to the relativistic Sedov-Taylor phase takes place at radius $\Rd$ (Eqn. \ref{eq:Rd}).
    With the values quoted above, and for $\eta~=~100$, the deceleration radius equals $\Rd = \Rst/\eta^2 \sim 2 \times 10^{-4}$~pc.
    In Section 3 we will consider various possibilities for placing the wind termination shock this close to the star.

\subsection{Confining pressure}

   The critical factor in bringing the wind termination shock closer to the star is the confining pressure of the surrounding medium.
   The pressure force constrains the expansion and size of the free-streaming wind region.
   The nature of this confining pressure depends on the circumstances.
   For instance, in the scenario described in Sect. \ref{sec-reverseshock} it is the thermal pressure in the bubble of shocked wind material,
   which in turn equals the ram pressure of the ISM associated with the overall expansion of the circumstellar bubble.
   For a rapidly moving star, it equals the ram pressure of the ISM associated with the stellar motion.
   One can generalize Eqn.~\ref{eq:Rshock1} to:
   \begin{equation}
   \label{eq:Rshock2}
      \Rw = \biggl(\frac{\mdotw \vw}{4 \pi \Pcf}\biggr)^{1/2}, \\
   \end{equation}
   with $\Pcf$ the confining pressure.

   If we stipulate that the GRB jet has to pass the wind termination shock before it enters the non-relativistic Sedov-Taylor expansion
   phase, this means that $\Rst~\geq~\Rw$.
   Combining Eqs.~\ref{eq:ST} and \ref{eq:Rshock2} yields:
   \begin{equation}
        \Egrb \geq \frac{\mdotw c^2}{\vw} \biggl(\frac{\mdotw \vw}{4 \pi \Pcf }\biggr)^{1/2} \; .
   \end{equation}
   Therefore, the confining pressure must satisfy:
   \begin{equation}
   \label{eqn:pcf1}
        \Pcf \geq \frac{c^4 \mdotw^3}{4 \pi \Egrb^2 \vw}.
   \end{equation}
   For the typical Wolf-Rayet wind values given in Sect. \ref{sec-blastwave} and for $E_{\rm GRB} = 2 \times 10^{52}$ erg,
   this gives a confining pressure equal to: $\Pcf~\simeq~1.8\times10^{-9}~{\rm dyne~cm^{-2}}$.

   If, on the other hand, we demand that the jet has to pass through the wind termination shock before it has started to decelerate,
   we get $\Rd~\geq~\Rw$.
   This would have the advantage that the entire afterglow is generated in the constant density medium, but this situation is much more difficult
   to bring about.
   From the relationship between $\Rst$ and $\Rd$ (Eqn.~\ref{eq:ST}) we find that the confining pressure now must satisfy:
   \begin{equation}
        \Pcf \geq \frac{\eta^4 c^4 \mdotw^3}{4 \pi \Egrb^2 \vw}.
   \end{equation}
   Since $\eta~>~100$, the confining pressure becomes extremely large: $\Pcf~\simeq~1~{\rm dyne~cm^{-2}}$ for typical parameters.
   Such a large confining pressure is almost impossible to create. For instance, the thermal pressure of a static gas is equal to:
   \begin{equation}
        P = 1.38\times10^{-16} \: n \:  T,
   \end{equation}
   with $n$ the particle density.
   So, for a particle density of 100\,000 particles cm$^{-2}$, which is already extremely high, we would need a temperature
   of approximately $10^{12}$~K.
   It is clearly not realistic to demand that the wind termination shock lies within the deceleration radius.
   In a similar fashion, if the confining pressure is a ram pressure,
   \begin{equation}
        P = \rho V^{2} = 1.67 \times 10^{-8} \: n \: \left(\frac{V}{1000 \: {\rm km/s}} \right)^2  \; ,
   \end{equation}
   it would require a large density in combination with a large velocity in order to achieve such a large value.
   Therefore, we will limit our discussion to the weaker condition (\ref{eqn:pcf1}), placing the termination shock radius well
   within the radius for non-relativistic Sedov-Taylor expansion.

    \begin{table}
    \centering
       \begin{tabular}{|l|r|r|r|}
          \hline
          \noalign{\smallskip}
              GRB  & redshift & $E_{iso}$  & density profile  \\
             name &          &($\times~10^{52}$~erg) &               \\
          \noalign{\smallskip}
          \hline
          \noalign{\smallskip}
           \object{GRB 970228}  &  0.695    &    2.68$^{(5)}$                     & w$^{(1)}$  \\
           \object{GRB 970508}  &  0.835    &    0.985$^{(5)}$                    & w$^{(1,3,4)}$ \\
           \object{GRB 980703}  &  0.966    &    8.09$^{(5)}$                     & u$^{(2)}$ \\
           \object{GRB 990123}  &  1.60     &    232$^{(5)}$                      & c$^{(1,2,3,4)}$  \\
           \object{GRB 990510}  &  1.619    &    21.9$^{(5)}$                     & c$^{(1,2,3,4)}$ \\
           \object{GRB 991208}  &  0.706    &    18.6$^{(5)}$                     & u$^{(2)}$, w$^{(3,4)}$ \\
           \object{GRB 991216}  &  1.02     &    7.83$^{(5)}$                     & w$^{(3)}$ \\
           \object{GRB 000301C} &  2.03     &    6.35$^{(5)}$                     & u$^{(3,4)}$  \\
           \object{GRB 000418}  &  1.118    &    11.6$^{(5)}$                     & w$^{(3,4)}$ \\
           \object{GRB 000926}  &  2.066    &    45.6$^{(5)}$                     & c$^{(3,4)}$ \\
           \object{GRB 010222}  &  1.477    &    154$^{(6)}$                      & w$^{(3)}$, u$^{(4)}$ \\
           \object{GRB 011121}  &  0.36     &    11$^{(6)}$                       & w$^{(4)}$ \\
           \object{GRB 020405}  &  0.69     &    7.37$^{(7)}$                     & w$^{(4)}$ \\
           \object{GRB 021004}  &  2.3      &    2.2$^{(8)}$                      & w$^{(4)}$ \\
           \object{GRB 021211}  &  1.01     &    0.61$^{(6)}$                     & w$^{(4)}$ \\
       \noalign{\smallskip}
       \hline
    \end{tabular}
    \caption{The redshift end isotropic energy for a number of GRBs as well as the density profile
    (w=wind, c=constant, u=undecided) of the circumstellar medium.
    If the density profile is marked as undecided, this means that both a wind and constant density profile are possible.
    Sources for the density profiles are:(1) Chevalier \& Li \cite{CL00}, (2) Panaitescu \& Kumar \cite{PK01}, (3) Panaitescu \& Kumar \cite{PK02} and (4)
    Chevalier et al \cite{CLF04}.
    The redshifts were taken from the table at: http://www.mpe.mpg.de/$\sim$jcg/ maintained by J.~Greiner.
    The values for the isotropic energy come from (5) Bloom et al. \cite{BFS01}; (6) Amati \cite{A03}, (7) Price et al.
    \cite{Petal03} and (8) Holland et al. \cite{Hetal03}.
    }
    \label{tab:GRBstat}
    \end{table}

\subsection{Observational indications}
\label{sec-afterglowmodels}

   The distance a GRB blast wave or jet will travel before it starts to decelerate
   depends on its initial energy $\Egrb$, and on the density
   of the surrounding matter. This means that a progenitor star that
   leaves a low-density wind, e.g. due to low metallicity,
   and which produces a high-energy GRB, might have a higher chance to
   show the GRB afterglow light curve for a constant density medium,
   since the jet can travel far before it starts to decelerate (Eqn. \ref{eq:Rd}).

\begin{figure}[!t]
    \centering
    \resizebox{\hsize}{!}{\includegraphics[width=0.95\textwidth]{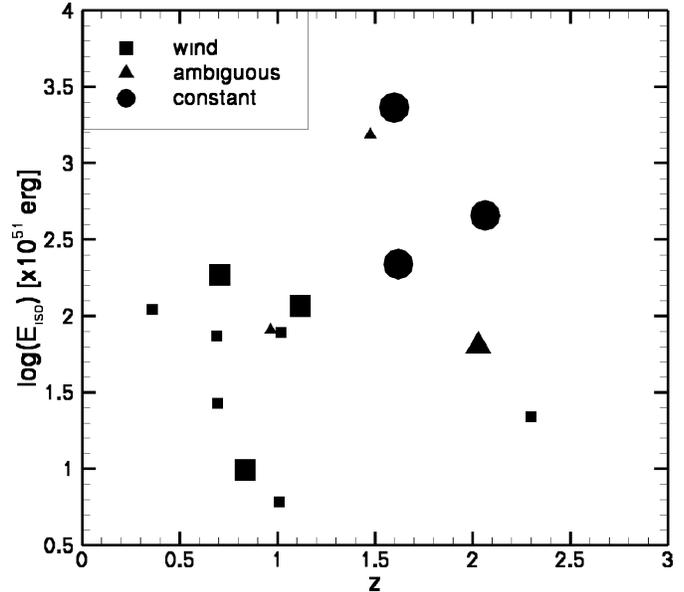}}
       \caption{Distribution of GRBs over the redshift-isotropic energy space (from Table~\ref{tab:GRBstat}).
       GRBs whose afterglow shows a constant density profile (circle) tend to have high energy and occur at larger redshifts than the
       afterglows which are formed in a free-streaming wind (square).
       The ones marked as triangles are ambiguous, which means that either different groups disagree on the nature of its afterglow, or no group
       has decided on its nature.
       If the symbol is large, several groups have reached the same conclusion.
       References for these afterglow models are: Chevalier \& Li (\cite{CL00}), Panaitescu \& Kumar (\cite{PK01}),
       Panaitescu \& Kumar (\cite{PK02}) and Chevalier et al. (\cite{CLF04}).
        }
        \label{fig:GRBstat}
    \end{figure}

   We have compiled a list of those GRBs for which indications about the circumstellar density profile of
   the CSM have been derived from the afterglow emission (Table \ref{tab:GRBstat}).
   Fig. \ref{fig:GRBstat} shows the location of these bursts
   in the isotropic energy-redshift plane.
   We find the bursts with clear indications of a constant density
   medium preferentially at high isotropic energy and high redshift,
   while the bursts with a free wind type CSM favor low energies
   and redshifts.

   This figure indicates that afterglows seem to follow a trend
   consistent with the considerations of this section.
   Furthermore, there seem to be more afterglows that show a wind-like
   CSM than afterglows with a constant-density environment.
   However, if afterglows with a constant density profile indeed occur
   on average at higher redshifts, the relative numbers of the two groups
   may change with the advent of new data from the SWIFT satellite.
   Also, in various individual cases, which are not considered here,
   a constant density profile is preferred (Piro et al. \cite{Petal05}).

   If the shocked wind does indeed provide the constant density medium observed in GRB afterglows, one would
   expect to see observational evidence that the GRB jet crosses the wind termination shock, as may have been observed in the afterglow of
   \object{GRB~050904} (Gendre et al. \cite{Getal06}). 
   However, the transition from the free-streaming wind to the constant density medium may not be clearly visible in the afterglow
   lightcurve (Nakar \& Granot, \cite{NG06}).

\section{Ways to get the wind termination shock close to the star}
\label{sec-windtermshock}

   The results derived in Sect. \ref{sec-reverseshock} provide us with a variety of ways in which the wind termination shock
   can be brought
   closer to the star. We have tested several scenarios, either analytically, numerically or both.
   For the hydrodynamical simulations we have used the ZEUS 3D code of Stone \& Norman (\cite{SN92}).

  \subsection{Low metallicity}
\label{sec-lowpr}
The winds of hot massive stars depends on their metallicity. 
For Gamma-Ray Burst progenitors, the mass loss rates might be lower than for Galactic Wolf-Rayet stars, since 
they may occur at low metallicity (Wijers \cite{W01}, Langer \& Norman \cite{LN06}). 
Such stars tend to have lower mass loss rates 
during their Wolf-Rayet phase (Crowther et al. \cite{Cetal02}, Vink \& de~Koter \cite{VK05}). 
Since the ram pressure of the wind depends linearly on the mass loss rate, this helps to bring the wind termination shock 
closer to the star.

From Eqn.~\ref{eq:Rshock1} we know that a change in mass loss rate of two orders of magnitude brings the wind termination 
shock only one order of magnitude closer to the star. 
Following Vink \& de~Koter (\cite{VK05}), Yoon \& Langer (\cite{YL05}) and Yoon, Langer \& Norman (\cite{YLN06}) give a metallicity dependence of the mass loss rate 
for Wolf-Rayet stars as:

\begin{equation}
\label{eq:metdep2}
{\rm log}\biggl(\frac{\mdotw}{\modot {\rm yr}^{-1}}\biggr) \propto 0.86 {\rm log}\biggl(\frac{Z}{{\rm Z_\odot}}\biggr).
\end{equation}

This implies that, in order to go from 10~pc to 0.1~pc, the mass loss rate would have to be {\bf four} orders of magnitude
lower. While it is not likely that many observed bursts have such low metallicities, values as low as 
$Z_{\odot}/100$ have been suggested based on observations (Chen et al. \cite{chen04}).

   In principle, one has to consider that the metallicity dependence of the
   mass loss rate applies to the main sequence phase of GRB progenitors as well.
   A star with a weak Wolf-Rayet wind had a weak main sequence wind before.
   This means that the thermal pressure in the hot bubble will be low (Eqn.~\ref{eq:Pbubble}), which puts the wind termination
   shock further away from the star.
   This effect decreases the overall effect of the weak Wolf-Rayet winds, but it is not strong: the inner radius
   of the wind bubble depends only on main sequence mechanical wind luminosity to the power of 0.2 (Eqn.~5).
   The dependence of the radius of the wind termination shock on the metallicity of the progenitor                           
    star was investigated by Eldridge et al. (\cite{EGDM05}). For stars at lower metallicity, the wind termination shock tends to lie closer to
    the star,but the dependence is weak.
  
   A benefit of low mass loss rates is the low density of the wind, which conforms to many of the afterglow
   lightcurve models (Chevalier et al. \cite{CLF04}).  
   This allows the GRB jet to penetrate further into the CSM before it decelerates (Eqn. {\ref{eq:ST}).
    This can explain why a signature of the wind termination shock does not appear in many GRB afterglow lightcurves:
    the jet is already past this point when the afterglow starts.

    \begin{table*}
       \caption{
               Assumed parameters for the three evolutionary stages of our $35~\modot$ model (Z=0.20).
               }
    \label{tab:windpar}
       \begin{tabular}{p{0.45\linewidth}rrrr}
          \hline
          \noalign{\smallskip}
              Phase & End of phase [\mbox{yr}] & $\mdotw$ [$\msy$] & $\vw$ [$\kms$] & $n_{\mathrm{photon}}$ [\mbox{s}$^{-1}$] \\
          \noalign{\smallskip}
          \hline
          \noalign{\smallskip}

           Main Sequence    & 4.745$\times10^6$ & 6.6$\times10^{-7}$ & 2\,671 & 3.4$\times10^{47}$ \\
           Red supergiant   & 5.098$\times10^6$ & 5.5$\times10^{-5}$ & 15     & 3.3$\times10^{41}$ \\
           Wolf-Rayet       & 5.295$\times10^6$ & 2.8$\times10^{-5}$ & 2\,140 & 3.3$\times10^{47}$ \\

          \noalign{\smallskip}
          \hline
       \end{tabular}
    \end{table*}

    \begin{figure}[!t]
    \centering
    \resizebox{\hsize}{!}{\includegraphics[width=0.95\textwidth]{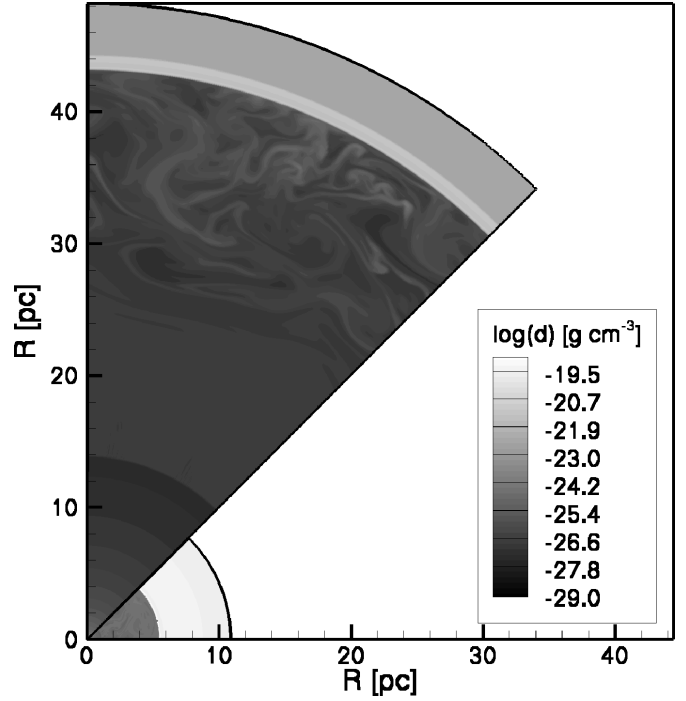}}

       \caption{The density of the gas around a 35 $\modot$ star with the density of the ISM set to $\rho_{0} = 10^{-22.5}~\gcm$ (upper segment)
       and $\rho_{0} = 10^{-19.5}~\gcm$ (lower segment).
        The high density of the ISM in the second segment limits the size of the bubble considerably, which increases the density in the bubble.
        Both segments show the same epoch, just before the star explodes as a supernova.
       }
          \label{fig:ismdens_A}
    \end{figure}

    \begin{figure}[!t]
    \centering
    \resizebox{\hsize}{!}{\includegraphics[width=0.95\textwidth]{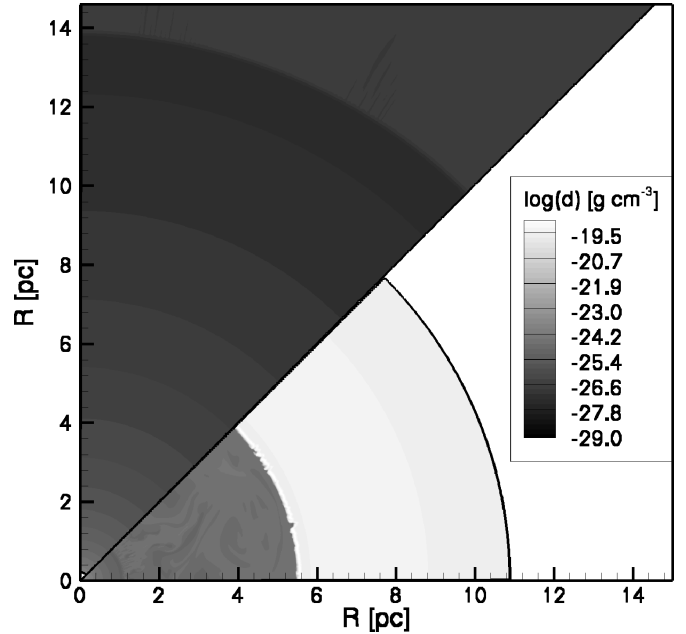}}
       \caption{A blow-up  of the central portion of Fig. \ref{fig:ismdens_A}.
       In the upper segment the wind termination shock can be seen at $R \sim 14$ pc.
       In the lower segment the termination shock is located at $R \sim 1$ pc.
       }
       \label{fig:ismdens}
    \end{figure}

\subsection{A high-density ISM}
\label{sec-ismdens}

    If the density of the ISM is increased, the whole wind-blown bubble becomes smaller,
    including the radius of the wind termination shock. A large range of densities in the ISM surrounding GRB progenitors is to be
    expected. Stars are formed inside molecular clouds, which means that the star may find itself in a high density region.
    Unfortunately, the dependence of the wind termination shock radius on the ISM density $\rho_{0}$ is not strong,
    $R_{\rm w}~\propto~\rho_{0}^{-3/10}$ (c.f. Eqn. \ref{eq:Rshock1}).
    This means that the density has to be very high in order to bring the wind termination shock as close to the star as we need,
    see for instance Eldridge et al. (\cite{EGDM05}).

    Figures \ref{fig:ismdens_A} and \ref{fig:ismdens} show the effect of a massive increase in the density of the ISM.
    Both segments of the figure show the density around a 35 $\modot$ star based on the star models calculated by Schaller et al. (\cite{SSMM92}).
    In one segment, the density has been set at $10^{-22.5}~\gcm$; in the other it is $10^{-19.5}~\gcm$.
    Both simulations were done using the method described in van Marle et al. (\cite{MLG04}, \cite{MLG05a}, \cite{MLG05b}).
    For the simulation with the low-density ISM the resolution is equal to 1000 radial grid points by 200 angular grid points.
    The high-density ISM simulation has a grid of 500 radial grid points by 200 angular grid points.
    The main-sequence and the start of the red supergiant phase are simulated in 1D, and the results are mapped onto the 2D grid.
    The end of the red supergiant phase, as well as the Wolf-Rayet phase, are simulated in two dimensions.

    The wind parameters for the three phases of the stars evolution can be found in Table~\ref{tab:windpar}.
    As can be expected from Eqn. \ref{eq:Rshock1}, a change in the ISM density by three orders of magnitude changes the location of the wind
    termination shock by approximately a factor of 10.
    To bring the wind termination shock even closer to the star would be difficult.
    To get a termination shock distance $R_{\rm w}~\sim~0.1$~pc one needs an ISM with a density of about $\rho_{0}~\sim~10^{-16}~\gcm$,
    which is rather unlikely.
    Again, as in Sect. \ref{sec-lowpr}, this scenario can only account for a few GRB progenitors.

    \begin{figure}[!t]
    \centering
    \resizebox{\hsize}{!}{\includegraphics[width=0.95\textwidth]{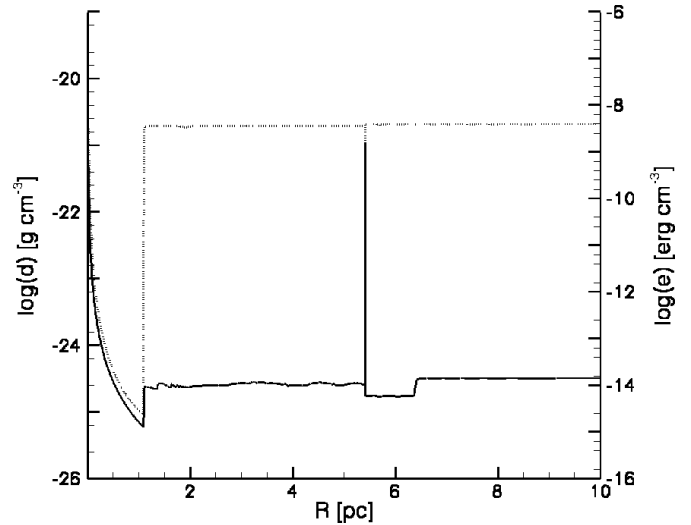}}
      \caption{The mass density (solid curve) and the internal energy density (dotted curve) of a circumstellar bubble that is
       expanding into a high-pressure ISM.
       Shown is the situation at a moment just before the star explodes as a supernova.
       Moving outward from the star (from left to right) we see the free-streaming Wolf-Rayet wind, the wind termination shock
       (at $R_{\rm w} \sim 1$ pc), the hot bubble filled
       with the shocked
        Wolf-Rayet wind material, the Wolf-Rayet wind driven shell (at $R \sim 5$ pc), the bubble of shocked main sequence wind material,
       and the ISM ($R \geq~6.5$ pc).
        }
          \label{fig:ismpres}
    \end{figure}

    \begin{figure}[!t]
    \centering
    \resizebox{\hsize}{!}{\includegraphics[width=0.95\textwidth]{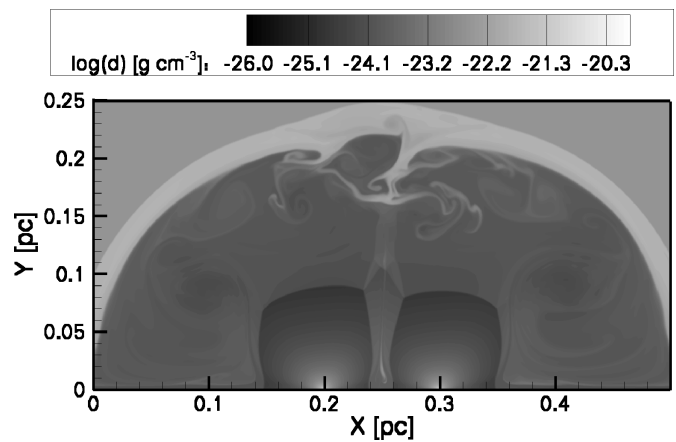}}
      \caption{The density of the gas around two massive stars during the main sequence.
       The CSM bubbles of the two stars have merged, creating a superbubble.
       The image corresponds to the situation approximately 800~years after the start of the evolution of the bubble.
       }
          \label{fig:2starsd}
    \end{figure}

    \begin{figure}[!t]
    \centering
    \resizebox{\hsize}{!}{\includegraphics[width=0.95\textwidth]{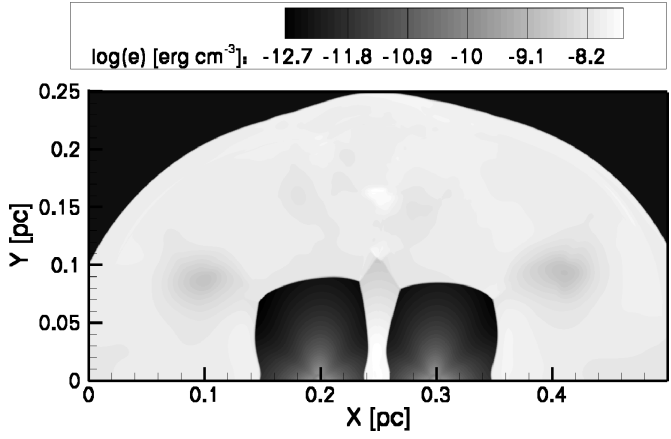}}
      \caption{The internal energy density of the CSM surrounding the two stars at the same epoch as in Fig. \ref{fig:2starsd}.
       The internal energy density is highest between the stars, where the stellar winds from both stars contribute to the heating of the gas.
       }
          \label{fig:2starse}
    \end{figure}

\subsection{A high-pressure ISM}
\label{sec-ismpres}
   The analytical solution described in Sect. \ref{sec-reverseshock} is only valid if the ISM is cold so that the circumstellar bubble
   expands supersonically with respect to the surrounding ISM.
   For a sufficiently high ISM temperature, the thermal pressure of the ISM will act as an extra confining force on the expanding shell.
   An example of such a situation is a circumstellar \ion{H}{II} region.
   If the star photo-ionizes the surrounding gas, the shell will expand more slowly.
   The expansion speed may even become subsonic. In that case the shell dissipates, and the circumstellar bubble will have a density
   discontinuity at the location where the shocked wind material meets the photo-ionized ISM.
   This effect has been described in van Marle et al. (\cite{MLG04}, \cite{MLG05a}, \cite{MLG05b}).

   A hot ISM is the explanation offered by Chevalier et al. (\cite{CLF04}) for the presence of matter with constant density close to the progenitor star.
   They describe the structure of a circumstellar bubble that has expanded into a low-density, extremely hot medium.
   In such a situation, the main sequence wind can never create a supersonically expanding shell.
   The expansion of the bubble remains subsonic, which means that the thermal pressure in the hot wind bubble must
   approximately equal the thermal pressure in the ISM.  Pressure balance between the wind ram pressure and the pressure
   in the ISM, $\rho_{\rm w} V_{\rm w}^{2} \sim P_{0}$, gives the radius of the wind termination shock as:
\begin{equation}
\label{eq:pressconf}
        R_{\rm w} \approx  \left( \frac{\mdotw \vw}{4 \pi P_{0}} \right)^{1/2} \; .
\end{equation}
   We have performed a simulation of a similar situation, the results of which are shown in Fig. \ref{fig:ismpres}.
   We used the same 35~$\modot$ star model that was used for the high-density simulation that was
   described in Sect. \ref{sec-ismdens}. The temperature of the ISM was set to $10^8$~K.
   The 1D radial grid has 10\,000 grid points.
   In Fig. \ref{fig:ismpres} the wind termination shock lies quite close to the star, once again at $R_{\rm w} \sim 1$ pc.
   The Wolf-Rayet wind-driven shell is still visible, even at the moment of supernova explosion.
   The reason is the high density and pressure of the main-sequence bubble.
   The pressure inside the shocked wind is high, since it has to be equal to the ISM pressure.  The density in this region
   is high because the whole bubble is very small: $\rho_{\rm b} \approx 10^{-25} \gcm$, which corresponds to
   a particle density $n_{\rm b} \sim 0.1 \; {\rm cm}^{-3}$, close to the value quoted by Chevalier et al. (\cite{CLF04}).
   This situation results in a slow expansion speed of the Wolf-Rayet shell.
   In a 2D or 3D simulation, the Wolf-Rayet shell would have been fractured by its collision with the Red Supergiant
   shell (Garc{\a'i}a-Segura et al. \cite{GML96}, \cite{GLM96}; van Marle et al. \cite{MLG05a}, \cite{MLG05b}) left by the progenitor
   in an earlier evolutionary phase.

   However, a circumstellar environment such as described here is difficult to create.
   Even if we consider a large association of massive stars that creates a {\em superbubble}, the merged wind-blown bubbles
   of a number of stars, it is almost impossible to increase the pressure to the level required to push the wind termination shock back to
   $R_{\rm w} \sim 0.1$ pc.
   Equation \ref{eq:pressconf} shows the problem quite clearly.
   If the wind termination shock has to be $\sim 10$ times closer to the star, the pressure in the surrounding ISM as to be a factor of 
   $\sim 100$ larger. This means that the total mechanical luminosity of the association that feeds the bubble has to be $\sim 10^5$ times larger
   than the typical mechanical luminosity of a single star, as the pressure in the bubble scales with the total mechanical luminosity as
   $P_{\rm b} \propto L_{\rm tot}^{2/5}$, c.f. Eqn. \ref{eq:Pbubble}.
   Effectively, this means that 100,000 massive stars have to release the kinetic energy of their wind into the bubble simultaneously.

   The only place where the gas pressure of the interstellar gas can reach the required level is in the interior of starburst regions. 
   There, stellar winds and supernova explosions add sufficient energy to the surrounding gas to reach these temperatures. 
   It should be noted, though, that the wind termination shock in Fig. \ref{fig:ismpres} is still more than one parsec away from the star. 
   To bring it closer by another order of magnitude would mean that the pressure of the surrounding gas has to be increased by two orders of
   magnitude.

\subsection{Colliding winds}
\label{sec-collwind}

   In reality the problem shown in Sect. \ref{sec-ismpres} may not be insurmountable provided two stars are sufficiently close together, for instance in a binary.
   In that case, the winds of the two stars collide directly, rather than simply feeding energy into the same superbubble.
   In the combined wind bubble the temperatures will rise very rapidly in regions where the winds collide.
   We will consider the binary scenario, in the following paragraph. In this paragraph we will consider the simpler case of two stationary stars, which are not gravitationally bound.

   Figs. \ref{fig:2starsd} and  \ref{fig:2starse} show what happens when two stars independently feed mechanical energy into the same bubble.
   The simulation was done in 2D on a cylindrical grid of 1000$\times$500 grid points.
   The two stars both have the same mass loss rate ($\dot{M} = 4\times10^{-6} \: \msy$) and wind velocity ($V_{\rm w} = 2000 \: \kms$).
   The two stellar winds have been simulated by filling two half spheres on the Z-axis with gas, moving away from central points (the stars)
   with a stellar wind density distribution (decreasing with the radius squared).
   The simulation includes radiative cooling, but the effect of photo-ionization is not considered.
   The winds never collide directly.
   Instead, each star creates its own bubble.
   When their respective main sequence shells hit each other, the bubbles merge.
   The region directly between the two stars is unrepresentative of the rest of the region, since the pressure is higher. 
   However, the difference is not enough to bring the termination shock close enough to the progenitor star, unless the GRB is aimed directly at
   the second star. 
   Still, if this were the case, the afterglow would not show a constant density medium, since the GRB jet would pass through the shocked wind material
   and run into the free-streaming wind of the second star.
   Should the winds of several stars collide directly, then the pressure in the bubble would locally be very high, which would indeed constrain the
   radius of the wind termination shock. 
   The problem remains that the GRB jet would pass quickly through the layers of shocked wind material and the afterglow lightcurve would most likely
   show a variety of density profiles.

    \begin{table}
       \caption{
               Wind parameters for the binary stars simulation shown in Fig.~\ref{fig:binary}.
               }
    \label{tab:binary}
       \begin{tabular}{p{0.45\linewidth}rr}
          \hline
          \noalign{\smallskip}
               & $\mdotw$ [$\msy$] & $\vw$ [$\kms$] \\
          \noalign{\smallskip}
          \hline
          \noalign{\smallskip}

           Star 1     & 1.0$\times10^{-5}$ & 1\,000  \\
           Star 2     & 1.0$\times10^{-4}$ & 10     \\

          \noalign{\smallskip}
          \hline
       \end{tabular}
    \end{table}

    \begin{figure}[!t]
    \centering
    \resizebox{\hsize}{!}{\includegraphics[width=0.95\textwidth]{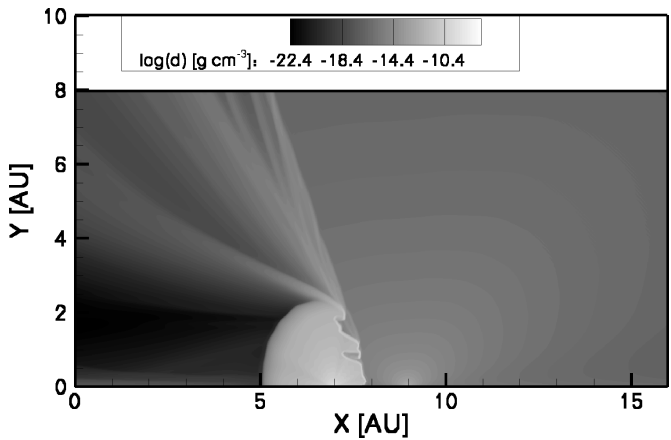}}
      \caption{The density of the gas around two stars with different wind parameters. The left star (at $X~=~7$~AU) has a high mass loss rate and
       a low wind velocity, while the right star (at $X~=~9$~AU) has a low mass loss rate but a high wind velocity, so that its ram pressure
       is the largest. The stronger wind (from the star on the right) dominates the CSM, while the weaker wind (on the left) is folded back,
       away from the companion star.
       }
          \label{fig:binary}
    \end{figure}

\subsection{A binary companion}
\label{sec-binary}

   The presence of a binary companion can bring the wind termination shock closer to the star
   For instance, in a Wolf-Rayet/O~star binary system, the Wolf-Rayet star, which has the smaller mass, will orbit the O-star.
   The Wolf-Rayet wind will be distributed over a larger volume, which decreases the density of the wind and its ram pressure.
   However, this effect is probably not very significant.
   If we are dealing with a close binary, the radius of the orbit is smaller than  $R_{\rm w}$ and the increase in volume is small.
   If we are dealing with a wide binary, the orbital motion will be so slow that the effect on the wind is no longer significant:
   to lowest order the Wolf-Rayet star can be treated as motionless.

   The wind emanating from the companion star is unlikely to be of any substantial influence: the ram pressure of a Wolf-Rayet star wind
   is much larger than that of the wind from nearly any other star.
   The wind from the companion will be pushed back in a bowshock, and its material will be constrained to a narrow layer
   in the orbital plane.
   This is demonstrated in Fig. \ref{fig:binary}.
   This is a simplified model of two stars in a binary system.
   The star on the left has a high mass loss rate, and a low wind velocity, like a red supergiant or luminous blue variable;
   the star on the right has a lower mass loss rate, but a high wind velocity, like a Wolf-Rayet star (Table~\ref{tab:binary}).
   The simulation is done in two dimensions, assuming a cylindrically symmetric situation.
   At the beginning of the simulation the grid is filled with ambient material at a constant density.
   The simulation includes radiative cooling, but the effect of photo-ionization is not considered.
   This simulation does not  take the orbital motion of the stars into account.
   It shows quite clearly how the weaker wind is being folded back by the strong Wolf-Rayet wind.
   An analytical solution pertaining to this situation was given by Canto et al. (\cite{CRW96}),
   while numerical simulations have been presented by Stevens et al. (\cite{SBP92}).

   The jet associated with the GRB is expected to move along the rotation axis of the progenitor star,
   which should be perpendicular to the orbital plane. Therefore, the companion star and its wind would have little or no
   influence on the GRB jet and the afterglow light curve.
   Of course, it is possible that a jet propagates through the wind of another star, which happens to be nearby.
   In that case, the density profile encountered by the GRB would depend entirely on the angle at which the jet hits the
   free-streaming wind. If it would meet the companion wind head-on, it would actually encounter a density that {\bf increases} as the
   jet penetrates deeper into the wind.

    \begin{table}
       \caption{
               Assumed parameters for the stellar wind in our bowshock model (Figs.~\ref{fig:bowshock1} to \ref{fig:bowshock3}).
               }
    \label{tab:bowshock}
       \begin{tabular}{p{0.45\linewidth}rr}
          \hline
          \noalign{\smallskip}
           phase  & $\mdotw$ [$\msy$] & $\vw$ [$\kms$] \\
          \noalign{\smallskip}
          \hline
          \noalign{\smallskip}

           Red Supergiant & 1.0$\times10^{-4}$ & 10        \\
           Wolf-Rayet     & 1.0$\times10^{-5}$ & 1\,000     \\

          \noalign{\smallskip}
          \hline
       \end{tabular}
    \end{table}

     \begin{figure}[!t]
     \centering
    \resizebox{\hsize}{!}{\includegraphics[width=0.95\textwidth]{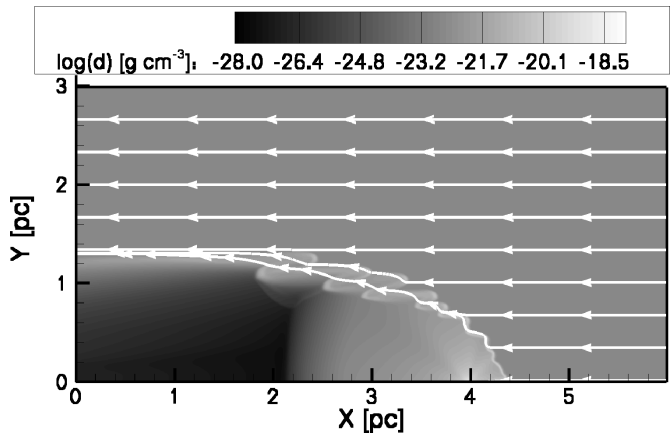}}

     \caption{The density of the gas around a moving red supergiant star.
       The velocity field in the ambient medium is shown as stream traces.
       The bowshock shell shows Kevin-Helmholtz instabilities.
       }
          \label{fig:bowshock1}
    \end{figure}

     \begin{figure}[!t]
     \centering
    \resizebox{\hsize}{!}{\includegraphics[width=0.95\textwidth]{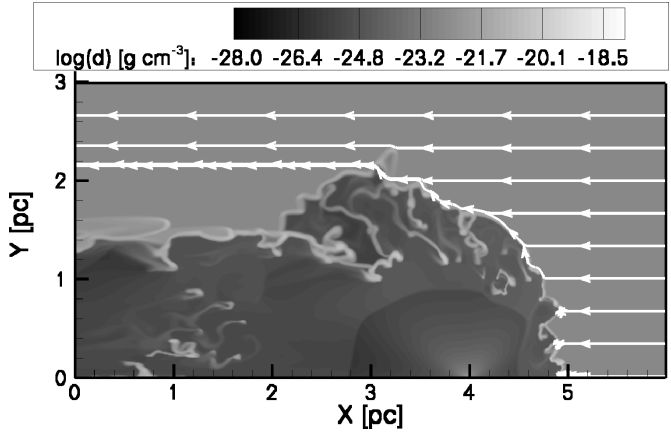}}
     \caption{Similar to Fig. \ref{fig:bowshock1}, but after the star has become a Wolf-Rayet star.
       The newly formed bowshock-shell system, which is sweeping up its predecessor, is highly unstable.
       }
          \label{fig:bowshock2}
    \end{figure}

    \begin{figure}[!t]
    \centering
    \resizebox{\hsize}{!}{\includegraphics[width=0.95\textwidth]{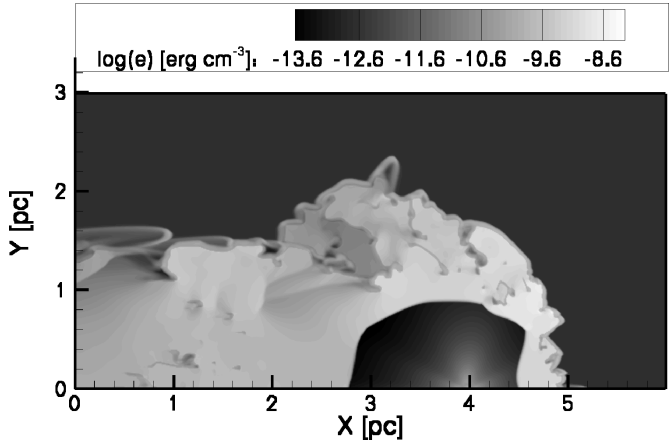}}
    \caption{The internal energy density at the same moment in time as in Fig. \ref{fig:bowshock2}.
       The wind termination shock is clearly visible, 0.5...1.25~pc away from the star, depending on the angle.
       }
          \label{fig:bowshock3}
    \end{figure}

\subsection{Stellar motion}
\label{sec-stellarmotion}

   If the progenitor star is moving rapidly through the interstellar medium, the situation becomes completely different.
   The equations in Sect. \ref{sec-reverseshock} apply when the star is motionless relative to the surrounding gas.
   However, if the star is moving at supersonic velocity with respect to the ISM, the collision between stellar wind and ISM will
   create a bowshock. This bowshock forms at the distance from the star $R_{\rm bs}$, where the ram pressure of the wind equals
   the ram pressure of the ISM in the frame where the star is at rest.
   The stand-off distance of the bowshock equals
   \begin{equation}
      R_{\rm bs} = \sqrt{\frac{\mdotw \vw}{4 \pi \rho_0 \: v_\star^2}},
   \label{eq:bowshock}
   \end{equation}
   where $v_\star$ is the velocity of the star (Chen et al. \cite{CWQ95} and Wilkin \cite{W96}).
   For a Wolf-Rayet wind, with a velocity of $2\,000~\kms$ and mass loss rate $3\times10^{-5}~\msy$,
   from a star that moves with a velocity of $80~\kms$ through an ambient medium with a density of $10^{-22.5}~\gcm$,
   this gives us a radius of the bowshock directly in front of the star of $R_{\rm bs}~\simeq~1.25$~pc.
   Generally, the radius of the bowshock depends on the angle between the position vector and the direction of motion.
   Similarly, the radius of the reverse shock is angle dependent, since the hot bubble of shocked wind material is not isobaric.
   The thermal pressure will be highest directly in front of the star, where the wind collides head-on with the ambient medium.

   Such a rapidly moving star can be the survivor of a close binary, where the companion has exploded as a supernova.
   In this example, the force of the supernova explosion broke up the binary, and the surviving star has moved away at approximately its original
   orbital velocity. Such runaway stars are not uncommon, but they do not always form a bowshock.
   If the star moves through a hot region of space (such as an \ion{H}{II} region, or the circumstellar bubble of another stars),
   its motion is subsonic and no bowshock is formed (Huthoff \& Kaper \cite{HK02}).
   During the early stages of its runaway phase, the star cannot form a bowshock, since it is moving through its own
   main-sequence wind bubble, where the temperature will be of the order of $10^7$~K, and the speed of sound is of the order of 300 km/s.
   How much time the star needs to escape its own bubble is extremely difficult to predict, as both the velocity of the star and the
   radius of the bubble can vary by at least an order of magnitude.  Typically, this will be of the order of $10^5...10^6$~years.
   Several Wolf-Rayet stars with bowshocks have been observed, such as \object{M1-67} around \object{WR~124} (van der Sluys \& Lamers \cite{SL03}) 
   and possibly \object{NGC~2359} around \object{HD~56925} (Chen et al. \cite{CWQ95})..

   To better understand the morphology of the circumstellar medium around a fast moving star, we have performed a two-dimensional
   hydrodynamical simulation.
   Since the stellar wind is assumed to be equally strong in all directions, it is possible to use a cylindrical grid,
   using the $\Phi$ axis as the symmetry axis. The $Z$-axis coincides with the direction of motion.
   This is similar to the approach used by Brighenti \& D'Ercole (\cite{BD97}) and Comer{\a'o}n \& Kaper (\cite{CK98}).
   Initially, the grid is filled with an ambient medium ($\rho_{0}=~10^{-22.5}~\gcm$), which travels along the Z-axis at a
   constant velocity ($v_\star~=~80\kms$). Matter flows in at one end of the grid,  and out at the other.
   The wind from the star is simulated by filling half a sphere on the Z-axis with material, which moves outward from a central point,
   and which has the density distribution
   of a stellar wind (as we did to model the binary system in Sect. \ref{sec-binary}).
   The grid resolution has to be rather high, in order to resolve the Kevin-Helmholtz instabilities in the shell (Brighenti \& D'Ercole \cite{BD97}).
   The stellar wind in our models follows two separate evolutionary phases: first the red supergiant phase, followed by a Wolf-Rayet phase.
   The wind parameters for these two phases can be found in Table~\ref{tab:bowshock}. Photo-ionization is not included in this simulation.

   In Figs. \ref{fig:bowshock1} through Fig. \ref{fig:bowshock3} we show the results.
   Figure \ref{fig:bowshock1} displays the density of the circumstellar medium during the red supergiant phase.
   The bowshock shell is clearly visible: it shows a series of so-called 'Cat's eyes', which are the result of a
   fully-developed Kevin-Helmholtz instability. These instabilities only occur if the ambient medium is cold.
   If the star moves through an \ion{H}{II} region, the instabilities disappear.
   In Figs. \ref{fig:bowshock2} and \ref{fig:bowshock3} the star and wind are in the Wolf-Rayet phase.
   The fast Wolf-Rayet wind has created a new shell, which sweeps up the old shell.
   It is extremely unstable.
   In Fig. \ref{fig:bowshock3}, which shows the internal energy density of the CSM, we can see the location of the wind termination shock.
   The pressure in the CSM is highest in front of the moving star, where the collision between wind and ambient medium is the most violent.
   The radius of the wind termination shock varies between 0.5 and 1.25~pc.

   Typically, one would expect the star to move in what was originally the orbital plane.
   This means that the GRB would be directed about $90^o$ away from the direction of motion,
   since the jets should move along the stellar rotation axis. The radius of the wind termination shock at that position
   angle is about~1~pc.

    \begin{figure}[!t]
    \centering
    \resizebox{\hsize}{!}{\includegraphics[width=0.95\textwidth]{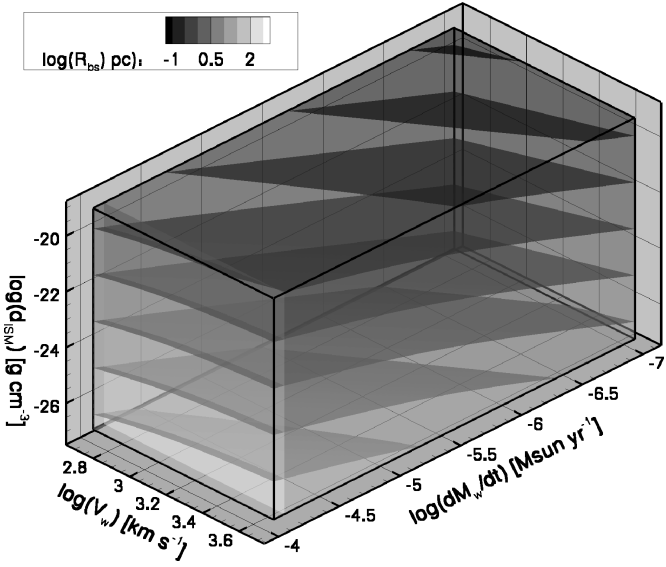}}
    \caption{The analytical solution for the radius of the reverse shock around a Wolf-Rayet star,
       depending on the parameters of the Wolf-Rayet wind and the density of the ISM.
       The radius of the reverse shock was calculated with Eqn. \ref{eq:Rshock1}.
       The Wolf-Rayet lifetime has been taken as constant and the wind parameters for the main sequence and red supergiant phase come from
       Table~\ref{tab:windpar}.
       The parameter space in this figure covers Wolf-Rayet wind velocities between 500 and 5000~$\kms$,
       mass loss rates between $10^{-7}$ and $1.0^{-4}~\msy$ and ISM densities ranging from
       $5~\times~10^{-4}$ to $5~\times~10^{4}$ particles per $\rm{cm}^{-3}$.
       The iso-surfaces in the plot show the planes with a constant radius of the wind termination shock.
       }
      \label{fig:reverseeshockradius}
    \end{figure}

    \begin{figure}[!t]
    \centering
    \resizebox{\hsize}{!}{\includegraphics[width=0.95\textwidth]{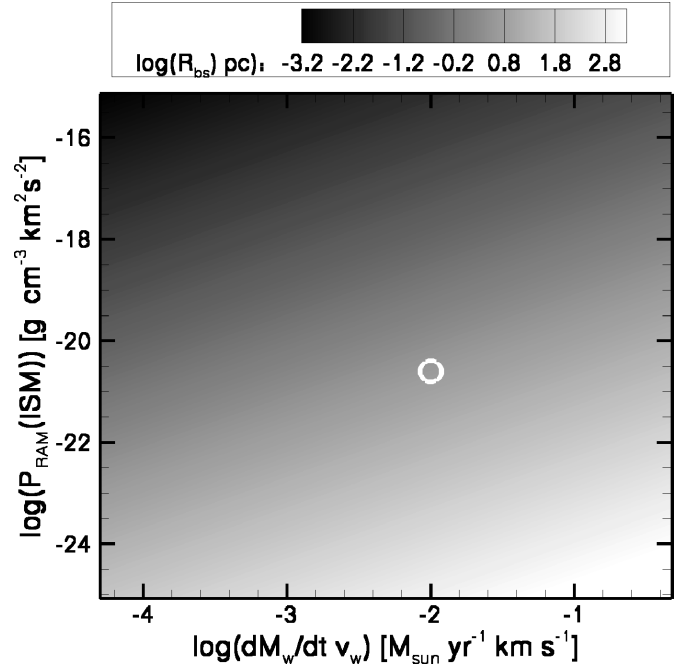}}
    \caption{The analytical solution for the radius of the bowshock in front of a Wolf-Rayet star, depending on the stellar wind parameters,
       the velocity of the star and the density of the ISM.
       On the horizontal axis we give the parameters of the stellar wind, on the vertical axis, the ram pressure of the ISM.
       The radius of the bowshock was calculated with Eqn. \ref{eq:bowshock}.
       The wind velocity, mass loss rate and ISM density cover the same intervals as in Fig. \ref{fig:reverseeshockradius}.
       The stellar velocity varies between 10 and 100~$\kms$.
       The circle shows the location of the numerically calculated model from Figs. \ref{fig:bowshock1} to \ref{fig:bowshock3}.
      }
      \label{fig:bowshockradius}
    \end{figure}


\section{Combining different scenarios}
\label{sec-combinations}

   It is possible that several of the mechanisms considered in Sect. \ref{sec-windtermshock} act together.
   However, not all combinations are possible.

\subsection{Low metallicity and high ISM density}

   A low ram pressure of the wind can be combined with any other mechanism, since it depends only on the internal structure of the star,
   while the other mechanisms rely on special external conditions such as the density or pressure in the ISM.
   The effect of combining different wind parameters and variations in the density of the ISM is shown in Fig. \ref{fig:reverseeshockradius}.
   Here we have taken a massive star and varied the wind velocity and mass loss rate during the Wolf-Rayet phase.
   As a basis we use the 35~$\modot$ star model from Table~\ref{tab:windpar}.
   The evolutionary time and the parameters for main-sequence and red supergiant wind are kept constant,
   but the parameters for the Wolf-Rayet wind are varied.
   The wind velocity of Wolf-Rayet stars varies between 500 and 5000~$\kms$, while the mass loss rate varies
   between $1.0^{-7}$ and $1.0^{-4}~\msy$ (c.f. Hamann et al. \cite{HKW95} and Koesterke \& Hamann \cite{KH95}).
   We have plotted the mass loss rate, the wind velocity and the density of the ambient medium in Fig. \ref{fig:reverseeshockradius}.
   The density of the surrounding ISM was varied between $5~\times~10^{-4}$ and $5~\times~10^{4}$ particles per $\rm{cm}^{-3}$.

   The results show a considerable variation in the location of the wind termination shock.
   As discussed above, the constant density inferred from some GRB afterglows is thought to be a signature of the shocked wind
   material. The quantitative value of this density can be calculated analytically, by dividing the total mass loss of the star over the
   spherical volume of the bubble. The outer edge of the bubble has a radius (Weaver et al. \cite{WCM77})
   \begin{equation}
   R = \biggl(\frac{250 \Lw}{308 \pi \rho_0} \biggr)^{1/5} t^{3/5} \; .
   \end{equation}
   Typically, the resulting particle density is less than 10~${\rm cm}^{-3}$ (see also Sect. \ref{sec-ismdens}, and the
   Figs. \ref{fig:ismdens_A},
   and \ref{fig:ismdens}),
   which is acceptable according to the observations (Stratta et al. \cite{Setal04}).
\subsection{Stellar motion in high density, and low metallicity}

   In Fig. \ref{fig:bowshockradius} we show how the radius of the bowshock varies with the stellar velocity $v_{\star}$ and the density of the
   interstellar medium $\rho_{0}$ in the combination $\rho_{0} v_{\star}^{2}$ (the ISM ram pressure),  and the wind parameters of the star
   in the combination $\dot{M} V_{\rm w}$, the absolute momentum discharge of the wind.
   The radius of the bowshock follows from Eqn. \ref{eq:bowshock}.
   This parameter study is simpler than the case of the stationary star, since the wind velocity and mass loss rate do not have to be plotted
   independently.
   The variations in wind parameters and ISM density are the same as those used to produce Fig. \ref{fig:reverseeshockradius}, while the stellar velocity
   varies between 10~$\kms$ and 100~$\kms$.
   Fig. \ref{fig:bowshockradius} shows that the radius of the bowshock changes considerably, depending on these parameters.
   The bowshock can be located closer to the star than the wind termination shock for a stationary star (see Fig. \ref{fig:reverseeshockradius}).

   In this figure we also show the location of the model that we simulated in the manner described in Sect. \ref{sec-stellarmotion}.
   The numerical result from the simulation is quite close to the analytical approximation.
   The dependence of the bowshock radius on the ISM density $\rho_{0}$ is much stronger than the density dependence of the wind termination
   shock radius for a stationary star, c.f. Eqns. \ref{eq:Rshock1} and \ref{eq:bowshock}, which makes this combination particularly effective.

    The main problem with the results of Fig. \ref{fig:bowshockradius} is the fact that, for extremely small bow-shock radii,
    the constant density is provided by the ISM, rather than by the shocked wind material.
    In those cases the density in the ISM is too high compared to the observations. These place a limit on the particle density of about
    10~${\rm cm}^{-3}$ (Stratta et al. \cite{Setal04}).
    This problem does not occur for GRB afterglows produced in a low-density shocked wind.

\subsection{Other combinations}

   Combining a high density with a high pressure in the ISM is nearly impossible, since a medium with a high thermal pressure will
   expand quickly (on its own hydrodynamical timescale), and the density will decrease rapidly.
   Actually, the high density would have little or no effect.
   If the pressure is as high as envisioned in Sect. \ref{sec-ismpres}, then the thermal pressure of the gas is much stronger than the
   ram pressure of a moving shell. This means that the thermal pressure is the only relevant confining force that acts on the shell,
   and the actual density of the ISM is no longer relevant.

   Binary companions can occur in any environment, irrespective of the density or pressure of the surrounding ISM.
   However, since the influence of a companion star on the location of the wind termination shock is limited at best, this makes little difference.

   Combining stellar motion with a high-pressure environment does not lead to a viable scenario, as the velocity of the star would automatically
   become subsonic, and the motion of the star would no longer be a relevant factor.
   In any case, the runaway star would have to live long enough for the star to leave its own circumstellar bubble,
   since the temperature of the shocked wind material would again make the stellar motion subsonic.

\section{Conclusions}
\label{sec-conclusion}

   Of the scenarios discussed, the most likely are: a Wolf-Rayet wind with low ram pressure, a high density
   ISM, and a moving progenitor star. It is the last possibility that
   can bring the wind termination shock extremely close to the star, as shown in Fig. \ref{fig:bowshockradius}.
   Observations show that the wind velocity and mass loss rate vary considerably from one Wolf-Rayet star to another.
   Since massive stars are thought to form within molecular clouds, a high density ISM is also possible.
   Fast moving Wolf-Rayet stars have been observed inside our galaxy.
   The combination of a weak Wolf-Rayet wind and a high ISM density can by itself be enough to obtain the desired result,
   but the choice of both parameters will have to be rather extreme.

   A high pressure ISM can produce the desired result as well, and probably accounts for some of the observed GRB afterglows.
   However, regions in the ISM with a sufficiently high temperature are rare.

   While many massive stars have binary companions, this by itself will hardly affect the 
   location of the wind termination shock, since the
   ram pressure associated with the wind from a Wolf-Rayet is much stronger than that of any possible companion.
   The Wolf-Rayet wind will therefore dominate the mass flow from the  binary system.
   Only if the binary companion were also a Wolf-Rayet star might its wind be competitive. 

   Only one such system has ever been found: \object{WR 20a} (Rauw et al. \cite{Retal04}), which contains two WN type stars. 
   However, since these two stars still have most of their hydrogen envelope, they are not yet 
   GRB progenitors. 
   Unlike most WN stars, these are main sequence stars with a very high mass loss rate.
   The orbital motion of a star in a binary system changes the properties of the stellar wind,
   but its effect is not enough to make a significant difference to the location of the wind termination shock.

  There is one possibility that has not been discussed so far:
  the constant density medium may not be in the hot bubble that drives the main sequence shell, but in the small bubble of shocked
  Wolf-Rayet wind material, which drives the Wolf-Rayet shell into the earlier Red Supergiant or Luminous Blue variable wind.
  For this to be true, the GRB would have to occur very early during the Wolf-Rayet phase.
  (Typically, a Wolf-Rayet wind driven shell moves at velocities of the order of 100 $\kms$, so it reaches 1~pc within 10\,000 years.)
  Apart from the extremely narrow time frame, there are several other problems with this scenario:
  During the early stages of its development, the layer of shocked Wolf-Rayet wind material is extremely thin
  ($\lesssim$~1~pc), so any afterglow generated in this medium would almost certainly show signs of either the transition
  between unshocked and shocked Wolf-Rayet wind, the Wolf-Rayet wind driven shell, or both.
  Second, the layer of shocked Wolf-Rayet wind, due to its thinness, rarely has a truly constant density, since both
  Rayleigh-Taylor instabilities of the Wolf-Rayet shell as well as internal turbulence in the shocked wind contribute to
  local density changes.

   The main problem of the scenarios discussed to bring the wind termination shock close to the star 
   is that the parameters of stellar mass loss or the conditions in the ISM have to be pushed to extremes.
   Therefore, none of the mechanisms described here seems, by itself, to account for the high
   percentage ($\sim 25$ \%) of the observed characteristics of
   GRB afterglows. However, the large number of possibilities and, in particular, possible combinations of the scenarios
   presented here, may explain why a significant fraction of GRB afterglows seem to occur in a constant-density medium.

   Generally speaking, outside influences like the motion of the star and the condition of the ISM can bring the termination shock to about 1~pc.
   In order to bring it any closer, the Wolf-Rayet wind has to be weak, which means that the progenitor star most likely had low metallicity.
   From our models of the circumstellar medium and the GRB afterglow models discussed in Sect. \ref{sec-afterglowmodels} we can conclude that
   the GRB afterglows generated in a constant density medium come from low metallicity stars with relatively strong GRBs
   (although it is an open question, as to whether the GRB itself was strong or the GRB jet had a narrow opening angle).

\begin{acknowledgements}
   This work was sponsored by the Stichting Nationale Computerfaciliteiten (National Computing Facilities Foundation, NCF), with financial support
   from the Nederlandse Organisatie voor Wetenschappelijk Onderzoek (Netherlands Organization for Scientific research, NWO).
   This research was done as part of the AstroHydro3D project:\\
   (http://www.strw.leidenuniv.nl/AstroHydro3D/) 
\end{acknowledgements}

\end{document}